\documentclass[twocolumn,floatfix,superscriptaddress,showpacs,showkeys,prl,aps]{revtex4}

\usepackage{amssymb}
\usepackage{amsmath}
\usepackage{amsfonts}
\usepackage{epsfig}
\usepackage{subfigure}
\usepackage{graphics}
\usepackage{float}
\usepackage{latexsym}

\begin{document}

\title{A Bayesian analysis of pentaquark signals from CLAS data}

%
%

\newcommand*{\ECOSSEG}{University of Glasgow, Glasgow G12 8QQ, United Kingdom}
\affiliation{\ECOSSEG}
\newcommand*{\ANL}{Argonne National Laboratory, Argonne, IL 60439}
\affiliation{\ANL}
\newcommand*{\ASU}{Arizona State University, Tempe, Arizona 85287-1504}
\affiliation{\ASU}
\newcommand*{\UCLA}{University of California at Los Angeles, Los Angeles, California  90095-1547}
\affiliation{\UCLA}
\newcommand*{\CSU}{California State University, Dominguez Hills, Carson, CA 90747}
\affiliation{\CSU}
\newcommand*{\CMU}{Carnegie Mellon University, Pittsburgh, Pennsylvania 15213}
\affiliation{\CMU}
\newcommand*{\CUA}{Catholic University of America, Washington, D.C. 20064}
\affiliation{\CUA}
\newcommand*{\SACLAY}{CEA-Saclay, Service de Physique Nucl\'eaire, 91191 Gif-sur-Yvette, France}
\affiliation{\SACLAY}
\newcommand*{\CNU}{Christopher Newport University, Newport News, Virginia 23606}
\affiliation{\CNU}
\newcommand*{\UCONN}{University of Connecticut, Storrs, Connecticut 06269}
\affiliation{\UCONN}
\newcommand*{\ECOSSEE}{Edinburgh University, Edinburgh EH9 3JZ, United Kingdom}
\affiliation{\ECOSSEE}
\newcommand*{\FU}{Fairfield University, Fairfield CT 06824}
\affiliation{\FU}
\newcommand*{\FIU}{Florida International University, Miami, Florida 33199}
\affiliation{\FIU}
\newcommand*{\FSU}{Florida State University, Tallahassee, Florida 32306}
\affiliation{\FSU}
\newcommand*{\GWU}{The George Washington University, Washington, DC 20052}
\affiliation{\GWU}
\newcommand*{\ISU}{Idaho State University, Pocatello, Idaho 83209}
\affiliation{\ISU}
\newcommand*{\INFNFR}{INFN, Laboratori Nazionali di Frascati, 00044 Frascati, Italy}
\affiliation{\INFNFR}
\newcommand*{\INFNGE}{INFN, Sezione di Genova, 16146 Genova, Italy}
\affiliation{\INFNGE}
\newcommand*{\ORSAY}{Institut de Physique Nucleaire ORSAY, Orsay, France}
\affiliation{\ORSAY}
\newcommand*{\ITEP}{Institute of Theoretical and Experimental Physics, Moscow, 117259, Russia}
\affiliation{\ITEP}
\newcommand*{\JMU}{James Madison University, Harrisonburg, Virginia 22807}
\affiliation{\JMU}
\newcommand*{\KYUNGPOOK}{Kyungpook National University, Daegu 702-701, South Korea}
\affiliation{\KYUNGPOOK}
\newcommand*{\UMASS}{University of Massachusetts, Amherst, Massachusetts  01003}
\affiliation{\UMASS}
\newcommand*{\MOSCOW}{Moscow State University, General Nuclear Physics Institute, 119899 Moscow, Russia}
\affiliation{\MOSCOW}
\newcommand*{\UNH}{University of New Hampshire, Durham, New Hampshire 03824-3568}
\affiliation{\UNH}
\newcommand*{\NSU}{Norfolk State University, Norfolk, Virginia 23504}
\affiliation{\NSU}
\newcommand*{\OHIOU}{Ohio University, Athens, Ohio  45701}
\affiliation{\OHIOU}
\newcommand*{\ODU}{Old Dominion University, Norfolk, Virginia 23529}
\affiliation{\ODU}
\newcommand*{\RPI}{Rensselaer Polytechnic Institute, Troy, New York 12180-3590}
\affiliation{\RPI}
\newcommand*{\RICE}{Rice University, Houston, Texas 77005-1892}
\affiliation{\RICE}
\newcommand*{\URICH}{University of Richmond, Richmond, Virginia 23173}
\affiliation{\URICH}
\newcommand*{\SCAROLINA}{University of South Carolina, Columbia, South Carolina 29208}
\affiliation{\SCAROLINA}
\newcommand*{\JLAB}{Thomas Jefferson National Accelerator Facility, Newport News, Virginia 23606}
\affiliation{\JLAB}
\newcommand*{\UNIONC}{Union College, Schenectady, NY 12308}
\affiliation{\UNIONC}
\newcommand*{\VT}{Virginia Polytechnic Institute and State University, Blacksburg, Virginia   24061-0435}
\affiliation{\VT}
\newcommand*{\VIRGINIA}{University of Virginia, Charlottesville, Virginia 22901}
\affiliation{\VIRGINIA}
\newcommand*{\WM}{College of William and Mary, Williamsburg, Virginia 23187-8795}
\affiliation{\WM}
\newcommand*{\YEREVAN}{Yerevan Physics Institute, 375036 Yerevan, Armenia}
\affiliation{\YEREVAN}
\newcommand*{\NOWUNH}{University of New Hampshire, Durham, New Hampshire 03824-3568}
\newcommand*{\NOWSACLAY}{CEA-Saclay, Service de Physique Nucl\'eaire, 91191 Gif-sur-Yvette, France}
\newcommand*{\NOWSCAROLINA}{University of South Carolina, Columbia, South Carolina 29208}
\newcommand*{\NOWUMASS}{University of Massachusetts, Amherst, Massachusetts  01003}
\newcommand*{\NOWMIT}{Massachusetts Institute of Technology, Cambridge, Massachusetts  02139-4307}
\newcommand*{\NOWECOSSEE}{Edinburgh University, Edinburgh EH9 3JZ, United Kingdom}
\newcommand*{\NOWGEISSEN}{Physikalisches Institut der Universitaet Giessen, 35392 Giessen, Germany}
\newcommand*{\NOWOHIOU}{Ohio University, Athens, Ohio  45701}
\newcommand*{\NOWECOSSEG}{University of Glasgow, Glasgow G12 8QQ, United Kingdom}

\author {D.G.~Ireland} 
\affiliation{\ECOSSEG}
\author {B.~McKinnon} 
\affiliation{\ECOSSEG}

\author {D.~Protopopescu} 
\affiliation{\ECOSSEG}

\author {P.~Ambrozewicz} 
\affiliation{\FIU}
\author {M.~Anghinolfi} 
\affiliation{\INFNGE}

\author {G.~Asryan} 
\affiliation{\YEREVAN}

\author {H.~Avakian} 
\affiliation{\JLAB}

\author {H.~Bagdasaryan} 
\affiliation{\ODU}
\author {N.~Baillie} 
\affiliation{\WM}

\author {J.P.~Ball} 
\affiliation{\ASU}
% ...

\collaboration{The CLAS Collaboration}
%
 
%

\date{\today}

\begin{abstract}

We examine the results of two measurements by the CLAS collaboration,
one of which claimed evidence for a $\Theta^{+}$ pentaquark, whilst
the other found no such evidence. The unique feature of these two
experiments was that they were performed with the same experimental
setup. Using a Bayesian analysis we find that the results of the two
experiments are in fact compatible with each other, but that the first
measurement did not contain sufficient information to determine
unambiguously the existence of a $\Theta^{+}$. Further, we suggest a
means by which the existence of a new candidate particle can be tested
in a rigorous manner.

\end{abstract}


\maketitle

The debate about the existence of the $S=+1$ $\Theta^+(1540)$ baryon
state is still going at this point in time in spite of results
from dedicated, high-luminosity measurements. One of these,
\cite{McKinnon:2006zv}, from the CLAS collaboration at the Thomas
Jefferson National Accelerator Facility used the reaction 
$\gamma d \rightarrow p K^{+} K^{-} n$. It showed convincing evidence
that production cross sections for such a state are nowhere near the
levels implied by an earlier CLAS measurement
\cite{Stepanyan:2003qr} of the same channel, which had seen a peak in
the $pK^{-}$ missing mass spectrum at
1.542 GeV/c$^{2}$ with a 5.2$\sigma$ statistical significance. The
salient point is that the work of Ref. \cite{McKinnon:2006zv} was a
dedicated, high luminosity repeat of Ref. \cite{Stepanyan:2003qr},
where the experimental running conditions were as similar as
practically possible.

In the whole history of $\Theta^+$ pentaquark searches, there were
several independent experiments that claimed to have found evidence,
whilst a similar number claimed to have found nothing. It is 
impractical to examine the results of all such experiments in a
consistent fashion, but the similarity of the two CLAS experiments
provides us with an ideal opportunity to investigate apparently
contradictory results.

One can examine in detail whether any discrepancy arose from the data
quality of the two experiments by making systematic tests on, for
example, the effects of different cuts. In the original work for both
measurements, however, parallel analyses were carried out to confirm
the final spectra, and different internal reviews verified the
correctness of the analysis procedures. We therefore assume that the
quality of the data in both the experiments was consistent, and that
the analyses of both experiments were carried out correctly. We
concentrate solely on the end-points of the analyses: namely, the
events passing all cuts, which contribute to missing mass spectra.

\begin{figure}[ht!]
\centering
\includegraphics[width=0.5\textwidth]{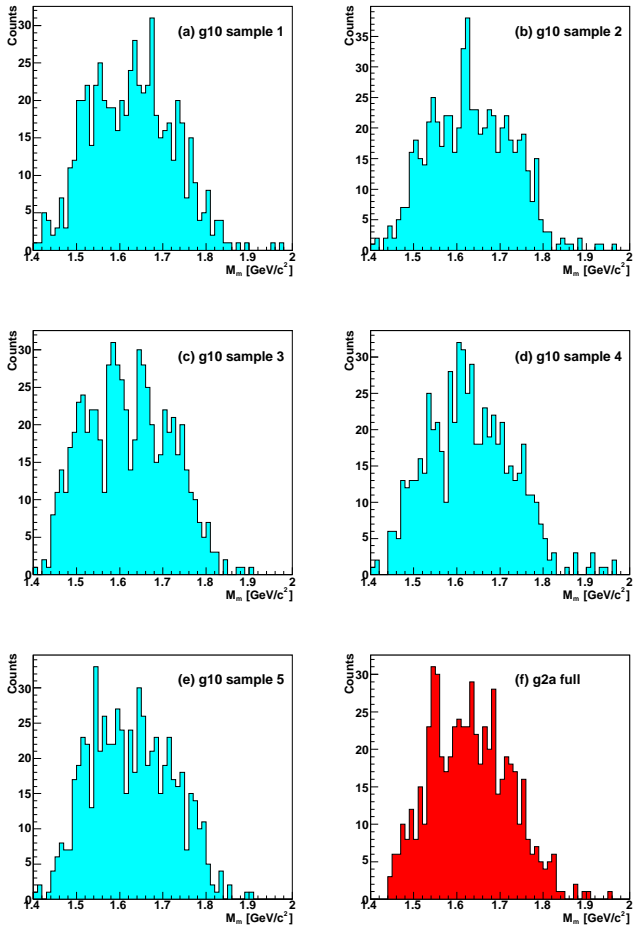}
\caption{\label{fig1}  (Color online) $pK^{-}$ missing mass spectra
from the five g10 subsamples and
the original g2a data. The data are sorted into bins of width 10
MeV/c$^{2}$. }
\end{figure}

To get a feel for the problem, we took the data set from Ref.
\cite{McKinnon:2006zv} (hereafter referred to as ``g10'' after the
CLAS running period in which the data was obtained) which had been
analyzed in exactly the same way as the data from
Ref. \cite{Stepanyan:2003qr} (hereafter referred to as ``g2a''). The
g10 data contained a factor of just under six more events, which could
be directly compared. The g10 data were then split into five
independent subsamples, each containing the same number of counts as
the g2a data set, and $pK^{-}$ missing mass spectra were
produced. These missing mass spectra would be where a $\Theta^+$ might
be expected to appear. The g10 subsample spectra are depicted in
figure \ref{fig1}a-e, and the g2a spectrum is depicted in figure
\ref{fig1}f.

Peak-like features appear in several of the g10 subsamples, but the
shapes are by no means consistent. As mentioned previously and in
keeping with current convention, the g2a result quoted a
``significance'' of about 5$\sigma$, which was similar to other
experiments claiming evidence of discovery. However, 5$\sigma$ means
that the probability that a feature is a fluctuation is of the order
of $10^{-6}$.  This is a very small number; it does not appear to
match the relative ease of generating peak-like features in the
subsample spectra. How do we quantify the intuitive feeling that the
odds of obtaining the observed g2a peak from fluctuations are not as
small as 1 in $10^{6}$ ?

In this letter we attempt to address this problem within a Bayesian
analysis framework, and to suggest an alternative means of quantifying
the evidence for discovery. 
What is specifically required is a quantitative comparison between two
hypotheses: ``the spectrum contains a peak'', and ``the spectrum does
not contain a peak''. One can model the shape of a spectrum as the
addition of simple functions, provided that they appear to describe
the shape of the spectrum reasonably well, and have plausible physical
origins (e.g. Gaussians for resolution effects, etc.). We refer to
these as ``data models'', to distinguish them from theoretical
models. The posterior probability that a data model ($M$) is true
given some observed data ($D$) is given by Bayes' theorem,
\begin{equation}
P\left( M \mid D \right) = 
\frac{P\left( D \mid M \right)P\left( M \right)}{P\left( D \right)},
\end{equation}
where $P\left( D \mid M \right)$ is the probability of the data being
observed given the model, and $P\left( M \right)$ represents the prior
probability of the model being correct. $P\left( D \right)$ is a
normalizing constant, which will cancel out in the ratio that compares
the posterior probabilities of two models.

Now the data model will depend on some parameters $\xi$, and the
posterior probability of these taking on specific values is
\begin{equation}
P\left( \xi \mid D, M \right) = 
\frac{P\left( D \mid \xi, M \right)P\left( \xi \mid M \right)}
{P\left( D \mid M \right)},
\label{eq:equ2}
\end{equation}
where $P\left( D \mid \xi, M \right)$ is the probability of the data
being observed given the model and its parameters, and $P\left( \xi
\mid M \right)$ is the prior probability of the parameters. Fitting
parameters to data is a matter of maximizing this posterior. The
quantity in the denominator of Eq.~(\ref{eq:equ2}) is known as the
\emph{evidence} for a model and is obtained by marginalizing
(integrating) over the parameters:
\begin{equation}
P\left( D \mid M \right) = 
\int{d\xi P\left( D \mid \xi, M \right)P\left( \xi \mid M \right)}.
\label{eq:evidence}
\end{equation}
Since the evidence is an integral over the model parameters, it
implicitly implements Occam's razor. Evidence ratios provide a balance
between favouring on the one hand the simpler model, and on the other
hand the model that better fits the data.

We construct two very simple data models of the missing mass spectra
obtained from experiment:

\begin{itemize}
\item Model $M_{0}$: The spectrum can be described by a $3^{rd}$ order
polynomial in the region of interest. This represents the assumption
that there is no new particle. A  $3^{rd}$ order polynomial was
employed in the original analysis to model the background shape. This
model depends on four parameters.
\item Model $M_{P}$: The spectrum can be described by a ``narrow''
Gaussian peak sitting atop a $3^{rd}$ order polynomial background in the
region of interest. ``Narrow'' in this case meaning that the width is
significantly less than the region of interest in the mass
spectrum. This model depends on seven parameters.
\end{itemize}

To compare the different models, a ratio of their probabilities in the
light of data can be formed:
\begin{equation}
R_{E} = 
\frac{P\left(M_{P}\mid D \right)}{P\left(M_{0}\mid D \right)} = 
\frac{P\left(D \mid M_{P} \right)}{P\left(D \mid M_{0} \right)}
\times
\frac{P\left(M_{P}\right)}{P\left(M_{0}\right)}\label{eq:ratio}, 
\end{equation}
where Bayes' theorem has been used to obtain the final
expression. This is the ratio of evidences for the models multiplied
by the ratio of prior probabilities of the models. If there is no
prior preference for either model, the final factor is unity, so the
ratio of model probabilities becomes a ratio of evidences. $R_{E}$ is
known as the ``Bayes' Factor'' or ``evidence ratio''.

It is computationally convenient and equivalent to examine the
logarithms of the evidence ratios:
\begin{equation}
\ln(R_{E}) = \ln P\left(D \mid M_{P} \right) - 
\ln P\left(D \mid M_{0} \right)\label{eq:lnratio}.
\end{equation}
Determining what value of $\ln(R_{E})$ to use in
deciding between data models is somewhat arbitrary, but Jeffreys
established \cite{Jeffreys:1961aa} a rough evidence 
scale versus 
written descriptors: $|\ln(R_{e})| < 1$ is {\it weak}, $1 < |\ln(R_{e})| <
2.5$ is {\it substantial}, $2.5 < |\ln(R_{e})| < 5$ is {\it strong} and
$|\ln(R_{e})| > 5$ is {\it decisive}. So model comparison is quantified by
$R_{E}$, and as constructed means that data favouring a data model
with a peak have positive $\ln(R_{e})$.

To evaluate evidences, we see from Eq.~(\ref{eq:evidence}) that an
integral over a likelihood $P\left( D \mid \xi, M \right)$ and a prior
$P\left( \xi \mid M \right)$ is required. We calculate the likelihood
by evaluating for each bin in a spectrum an ``ideal'' number of counts,
$S_{i}(\xi)$, for a given set of parameters. The
probability of this being correct given the measured counts $n_{i}$ is
calculated
using a Poisson distribution. The total likelihood is then a product
of these probabilities for each bin:
\begin{equation}
P\left( D \mid \xi, M \right) = 
\prod_{i}\frac{S_{i}^{n_{i}}
\exp\left(-S_{i}\right)}{n_{i}!}.
\label{eq:likelihood}
\end{equation}

Here, the prior probability is constructed by assuming no initial
correlations between parameters, so it is simply a product of priors
for each separate parameter. We assume that each prior is a uniform
distribution between a lower and upper limit since this represents the
least initial bias. The prior parameter ranges were established by
performing an initial fit and setting the limits to be $\pm$50\% of
the values found. This resulted in a large flexibility in the shapes
of both background and peak.

To perform the integrations over the many parameters in the models, we
utilized the technique of ``nested sampling'' developed by Skilling
\cite{nested,siv06}. Essentially, this is a Monte Carlo integration
method developed specifically for Bayesian data analysis. We refer the
reader to the original reference for details, and to
Ref.~\cite{Mukherjee:2005wg} for an example application.

\begin{figure}[htp!]
\centering
\includegraphics[width=0.5\textwidth]{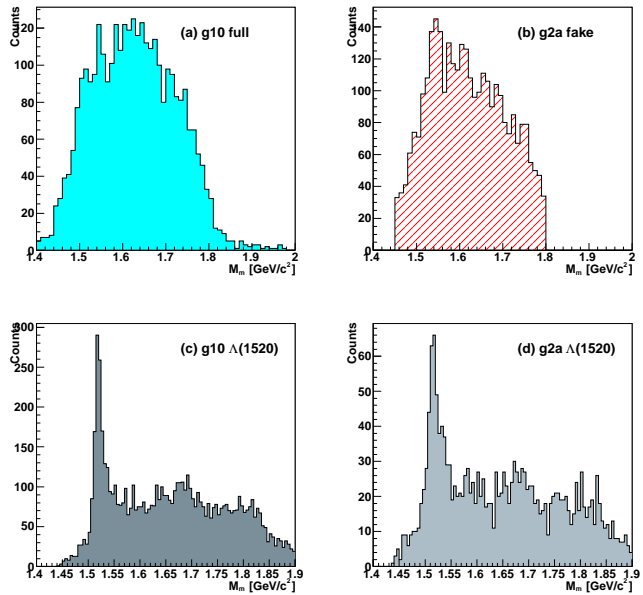}
\caption{\label{fig2}  (Color online) Missing mass histograms for
$\Theta^+$ from a) g10, b) fake, and $\Lambda(1520)$ from c) g10, and
d) g2a data. The data in a) and b) are sorted into bins of width 10
MeV/c$^{2}$, and the bins in c) and d) have width 5 MeV/c$^{2}$.}
\end{figure}

We applied the model comparison framework to all the spectra shown in
figure (\ref{fig1}). In addition we analyzed the spectra shown in
figure (\ref{fig2}), which consisted of: (a) the full g10 spectrum; (b)
a ``fake'' spectrum, constructed by sampling from a combination of
signal and background functions in the data model with the
peak ($M_{P}$), which had the same signal-to-background ratio as
the g2a spectrum. This was done to show what the results of this
analysis would have been, had a resonance been there; (c) and (d)
$pK^{-}$ missing mass spectra from the g2a and g10 data sets, but
showing the $\Lambda(1520)$ signal, in order to test how the technique
fared for the case of a well-established particle.

\begin{figure}[t!]
\centering
\includegraphics[width=0.48\textwidth]{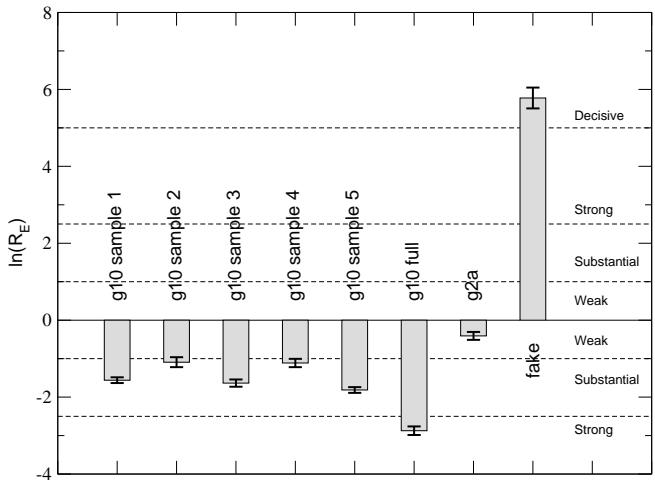}
\caption{\label{Fig3}  Graphical representation of the
values of evidence ratios    from table \ref{EIRNS}, on  a   logarithmic
scale. The horizontal lines correspond to the limits of the
regions associated with the different descriptors of the Jeffreys scale.}
\end{figure}

The results are quoted in table (\ref{EIRNS}), and displayed
graphically in figure \ref{Fig3}. We omit the results for the
$\Lambda(1520)$ from the figure, as they would render the scale
unusable. To estimate the uncertainty in the Monte Carlo integrals, we
ran at least 20 independent calculations for each spectrum
analysed. The errors listed in the table represent the standard error
of the samples.

\begin{center}
\begin{table}[b!]
\begin{tabular}{|l|rcl|}
\hline\hline
Data sample & & $\ln(R_{E})$  & \\
\hline
 g10 sample 1 & -1.56 & $\pm$ &  0.07 \\
 g10 sample 2 & -1.09 & $\pm$ &  0.13 \\
 g10 sample 3 & -1.64 & $\pm$ &  0.09 \\
 g10 sample 4 & -1.11 & $\pm$ &  0.11 \\
 g10 sample 5 & -1.82 & $\pm$ &  0.07 \\
\hline
 g10 full & -2.87 & $\pm$ &  0.11 \\
\hline
 g2a & -0.41 & $\pm$ &  0.10 \\
 fake & 5.78 & $\pm$ &  0.27 \\
\hline 
g2a $\Lambda(1520)$ &   96.70 & $\pm$ & 0.70  \\
g10 $\Lambda(1520)$ &  549.12 & $\pm$ & 2.17  \\
\hline\hline
\end{tabular}
\caption{\label{EIRNS} Evidence ratios. Calculations are done by
nested sampling, hence the need to include standard errors.}
\end{table}
\end{center}

With the splitting of the g10 data set, we have shown (figure
\ref{fig1}) the relative ease with which one can obtain a peak-like
feature, given a small number of events. The evidence ratios 
calculated for the individual subsamples in g10 generally suggest a 
bias against a peak, which perhaps mirrors an intuitive feeling about 
how significant such features really are. However, two of the five
subsamples (2 and 4) are compatible with the ``weak" category, meaning
that the results are essentially inconclusive. Whilst the g2a result
is more of an outlier, it also falls in the weak category and is
inconclusive; the results of the two measurements are therefore
compatible with each other. 

The $\ln(R_{E})$ value for g2a (-0.408) indicates weak evidence in
favour of the data model without a peak in the spectrum. What this
means is that whilst a data model including a peak gives a better fit
by eye to the spectrum, it does not compensate for having had to
introduce additional parameters for the peak. This is Occam's razor in
action; simpler models are preferable unless more complex models do
much better. One must be careful what to conclude from the g2a
spectrum, however, since the evidence ratio does {\it not}
conclusively rule out a peak; it is simply inconclusive. 

We now turn to the question of whether the g10 experiment could
conclusively discriminate between the two possibilities. The log of
the evidence ratio for the full g10 spectrum is -2.9. This makes it
{\it strong} evidence against a peak in the spectrum. Another way of
looking at this is that with this evidence ratio, the odds against a
peak in this spectrum are about 17 to 1. Whilst this cannot completely
rule out a discovery, another measurement of this channel is probably
not necessary. By comparison, the odds in favour of a peak in the fake
spectrum are about 320 to 1, meaning that had a signal really been
there in g10, the experimental result would have been {\it decisive}.

The study of the $\Lambda(1520)$ shows that when a resonance is there,
this method picks it out rather readily, with both g2a and g10 data
sets yielding a {\it decisive} result. We take this as a positive test
that our method works.

In summary, we have applied a Bayesian model comparison method to
analyzing the missing mass spectra produced in pentaquark
searches. This has been used to study the relationship between the
results of two CLAS measurements, which were taken under almost
identical conditions. We have shown that there is no conflict between
the results of the two experiments, and that the low number of counts
in the first experiment resulted in an ambiguous signal. Furthermore
we have shown that the g10 result shows strong evidence against the
discovery of a pentaquark in this channel. More generally, this method
could be applied to any data set where a search for a new state has
been carried out, and can provide a quantitative measure with which to
judge whether or not a result represents a discovery.

\subsection{Acknowledgments}


We would like to thank  G. Woan  (Glasgow) for  useful
discussions.  We would also like to thank the staff of the Accelerator
and  Physics  Divisions  at Jefferson Lab  who   made  the experiments
possible.   Acknowledgments for  the support  of  these experiments go
also to the Italian Istituto Nazionale de  Fisica Nucleare, the French
Centre  National  de la  Recherche   Scientifique and Commissariat \`a
l'Energie Atomique, the Korea Research Foundation, the U.S. Department
of   Energy     and  the  National   Science    Foundation,    and the
U.K. Engineering and Physical Science Research Council.  The Jefferson
Science Associates   (JSA) operates   the  Thomas Jefferson   National
Accelerator Facility for the  United States Department of Energy under
contract DE-AC05-06OR23177.


\end{document}